\documentclass[runningheads]{llncs}
\usepackage[T1]{fontenc}
\usepackage{graphicx,verbatim}
\usepackage{subfigure}
\usepackage{url}
\usepackage{color,soul}
\usepackage{amssymb}
\usepackage{cleveref}
\usepackage{array}
\usepackage{enumitem} 

\usepackage{multirow}
\usepackage{makecell}
\crefname{figure}{Fig.}{Figs.}  
\Crefname{figure}{Figure}{Figures}  
\crefname{table}{Tab.}{Tabs.}  
\Crefname{table}{Table}{Tables}  

\begin{document}
\title{Markerless Tracking-Based Registration for Medical Image Motion Correction}
\titlerunning{Markerless Tracking-Based Registration for Motion Correction}
%
\author{Luisa Neubig\inst{1}\and
Deirdre Larsen\inst{2} \and
Takeshi Ikuma\inst{3} \and
Markus Kopp\inst{4}\and
Melda Kunduk\inst{5} \and
Andreas M. Kist\inst{1}} 

%
\authorrunning{Neubig et al.}
%
\institute{Department Artificial Intelligence in Biomedical Engineering, FAU Erlangen-Nürnberg, Erlangen, Germany\\
\and
Department of Communication Sciences and Disorders, East Carolina University, Greenville, NC, USA\\
\and
Department of Otolaryngology–Head and Neck Surgery, Louisiana State University Health Sciences Center, New Orleans, LA, USA\\
\and
Department of Radiology, FAU Erlangen-Nürnberg, Erlangen, Germany\\
\and
Department of Communication Sciences and Disorders, Louisiana State University, Baton Rouge, LA, USA\\
\email{luisa.e.neubig@fau.de}}



\maketitle              
\begin{abstract}

Our study focuses on isolating swallowing dynamics from interfering patient motion in videofluoroscopy, an X-ray technique that records patients swallowing a radiopaque bolus. These recordings capture multiple motion sources, including head movement, anatomical displacements, and bolus transit. To enable precise analysis of swallowing physiology, we aim to eliminate distracting motion, particularly head movement, while preserving essential swallowing-related dynamics. Optical flow methods fail due to artifacts like flickering and instability, making them unreliable for distinguishing different motion groups. We evaluated markerless tracking approaches (CoTracker, PIPs++, TAP-Net) and quantified tracking accuracy in key medical regions of interest. Our findings show that even sparse tracking points generate morphing displacement fields that outperform leading registration methods such as ANTs, LDDMM, and VoxelMorph. To compare all approaches, we assessed performance using MSE and SSIM metrics post-registration. We introduce a novel motion correction pipeline that effectively removes disruptive motion while preserving swallowing dynamics and surpassing competitive registration techniques.

\end{abstract}

\section{Introduction}
Motion analysis plays a critical role in medical imaging, particularly in the evaluation of dynamic processes such as swallowing~\cite{kang2010influence}.
Videofluoroscopy Swallowing Studies (VFSS) capture high-temporal-resolution (up to 30 fps) X-ray recordings of patients swallowing a radiopaque bolus to assess swallowing function~\cite{martin2020best}.
However, these recordings contain multiple overlapping sources of motion, including bolus flow, head movement, and anatomical structure shifts. This makes it difficult to focus and isolate clinically relevant swallowing dynamics. Accurately correcting for these confounding motions is crucial for precise physiological analysis. To achieve this, we must (i) assess global motion artifacts, (ii) identify distinct moving structures, and (iii) register relevant features to compensate for unwanted motion.
In this study, we address these challenges by isolating unwanted motion, such as head movement, while preserving anatomical motions and accurate depiction of the bolus suitable for downstream quantitative analysis. By leveraging advanced tracking algorithms, we compute velocity and displacement fields to effectively suppress large, distracting shifts. We show that our proposed method works on rather still and more challenging data, as well as generalizes to data provided by other hospitals.

\section{Related Works}

\subsection{Optical Flow-based Methods in Medical Imaging}
Optical flow methods are widely used in medical image analysis to estimate motion from pixel displacements and enhance computer vision tasks like segmentation and image registration. Xue et al.~\cite{motionsegmentation} propose a novel segmentation method for echocardiography that effectively utilizes motion information by accurately predicting the optical flow. This information and initial segmentation guesses result in a motion-enhanced segmentation module for final segmentation. 
Suji et al. analyzed how optical flow methods (i.e., Farneback, Horn-Schunck and Lucas-Kanade) can improve motion-based segmentation of lung nodules in thin-sliced CT scans~\cite{lungnodulesegmentation}. In~\cite{us-tracking}, motion information is used to discriminate between healthy and affected patients. They use an optical flow-based method to quantitatively analyze the deformation of the right diaphragm in ultrasound imaging to track respiratory motion.

\subsection{Image Registration in Medical Imaging}
Image registration enables the matching of medical images across time, modalities, or patients, as well as multimodal fusion to improve diagnosis and treatment. Deformable image registration for image-guided adaptive radiotherapy has issues in low-contrast regions. Meng et al.~\cite{aanetworks} address this problem by integrating a finite element method (FEM) with the 'demon' algorithm and refining a tetrahedral mesh to improve displacement accuracy and reduce registration errors. A common issue in medical image registration is that labeling corresponding features is a very time-consuming task, which is why image registration is preferably performed unsupervised. \cite{sam_reg} overcomes the labeling of corresponding areas by using foundation segmentation models, such as Segment Anything (SAM), to define ROI correspondences for registration.

\subsection{Tracking in Videofluoroscopies}
Previous studies on VFSS have focused on the dynamics of anatomical landmarks, assuming a motionless patient. However, we are not aware of any study that investigated or corrected for patient motion during VFSS recordings. On rare occasions, the community acknowledges this issue: ~\cite{hyoidtracking} draws attention to the importance of tracking the movement of the hyoid bone during VFSS. We believe that quantitative measurements such as in~\cite{hyoidtracking}, and~\cite{hyoidtracking2} or tracking of cervical vertebrae such as in~\cite{cervtracking} can benefit from compensating for the patient's head movement as we propose in our study.

\section{Methods}

\subsection{Dataset}
We created two datasets termed~\emph{motion}~\includegraphics[width=0.025\textwidth]{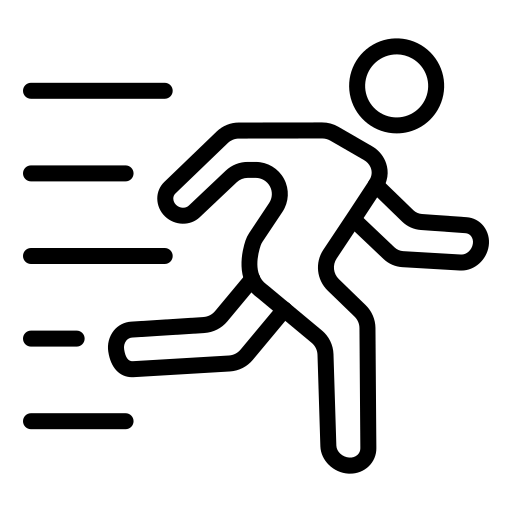} and~\emph{non-motion}~\includegraphics[width=0.025\textwidth]{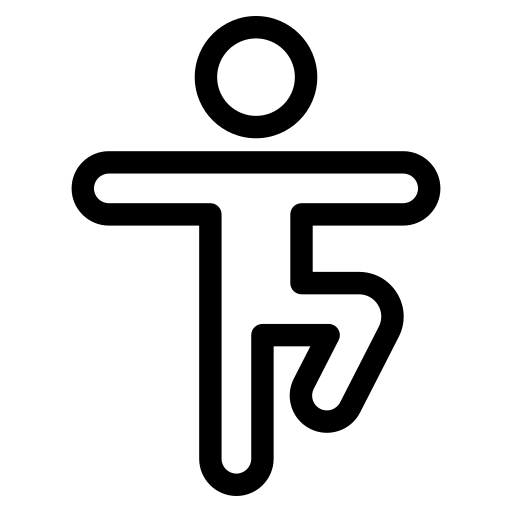} containing each of 10 manually selected VFSS recordings with a single swallowing event. The VFSS recordings were acquired at an American hospital (anonymized) and in accordance with the granted IRB~(\#IRBAM-anonymized). The videos have different lengths, ranging from approximately 30 to 170~frames. All videos have a temporal resolution of 30~fps. All frames were resized to a uniform resolution of $256\times256$~px. For validation purposes, we rely on example data from a European hospital (anonymized). This data is in many ways dissimilar to the American dataset, i.e., the temporal resolution is only 7.5~fps and the aperture is rectangular instead of circular.

\subsection{Assessing Optical Flow using Markerless Tracking}

We first analyzed multiple optical flow methods including classical approaches (Lucas-Kanade Method, Farneback) and Deep Learning-based methods (RAFT, SpyNet, FlowFormer, PWCNet, UnFlow). Next to these methods, we tested three different markerless tracking algorithms (CoTracker, PIPs++ and TAP-Net) to determine correspondences. All experiments were performed on an NVIDIA GeForce RTX 4090 with 24GB RAM. Additionally, we experimented with different grid sampling strategies to assess their impact on tracking performance and adjusted grid sizes to meet the GPU RAM limitations.

\begin{itemize}[align=right,label={},labelsep=2pt,labelwidth=1em,leftmargin=0pt]

\item 
\noindent\textbf{CoTracker}
utilizes a query-based mechanism, allowing for adaptive point selection and refinement across frames~\cite{cotracker}. We employed CoTrackerPredictor3 in its default offline configuration with the provided pre-trained model~\cite{cotracker3_facebook}.

\item 

\textbf{PIPs++}
uses a sliding window to update predictions iteratively and maintain trajectory consistency~\cite{pips}. 
We utilized PIPs++ in its default configuration and leveraged the provided pre-trained model~\cite{pips2_github}.

\item
\textbf{TAP-Net}
follows a two-stage approach to estimate trajectories accurately over time.
In the matching stage, TAP-Net independently searches for a suitable candidate point match for the query point across all frames, leveraging local and global spatial information. 
We use the provided pre-trained BootsTAPIR model for our analysis~\cite{bootstapir2024}.
\end{itemize}

\subsection{Motion Correction using Velocity Flow Fields}
We stabilize the patient during swallowing by correcting motion through image registration using the predicted velocity field. To evaluate our approach, we compare it with cross-domain state-of-the-art (SOTA) registration algorithms, as no existing methods are specifically designed for VFSS. The comparison includes Large Deformation Diffeomorphic Metric Mapping (LDDMM)~\cite{lddmm}, Advanced Normalization Tools (ANTs)~\cite{ants}, and VoxelMorph~\cite{voxelmorph}.

\subsubsection{LDDMMs}
formulate registration as finding a smooth, time-varying velocity field $\{v_t\}_{t \in [0,T]}$ whose flow integrates to a diffeomorphism. Specifically, we define the transformation $\phi_t : \Omega \to \Omega$ by
\begin{equation}
\frac{\partial \phi_t(x)}{\partial t} = v_t(\phi_t(x)), 
\quad \phi_0(x) = x,
\end{equation}
so that $\phi_1$ is the final deformation at $t=1$. The objective function typically balances a regularization on the velocity field (encouraging smoothness) and an image similarity term:
\begin{equation}
\min_{v_t} \int_0^1 \mathcal{R}(v_t)\, \mathrm{d}t 
\;+\; \mathcal{D}\bigl(I_0 \circ \phi_1^{-1}, I_1\bigr),
\end{equation}
where $\mathcal{R}$ denotes a regularization functional (e.g., $\| \nabla v_t \|^2$) and $\mathcal{D}$ is a measure of image dissimilarity (e.g., squared difference or mutual information).

\subsubsection{ANTs} implements a symmetric diffeomorphic registration (SyN), jointly estimating forward and inverse transformations in a single optimization framework. A velocity field is used to generate diffeomorphic transformations via the exponential map, ensuring invertibility. The energy functional typically combines an image similarity term, $\mathcal{D}$, and a smoothness regularizer, $\mathcal{R}$, on the deformation:
\begin{equation}
\phi^* = \arg \min_{\phi} \Bigl[
\mathcal{D}\bigl(I_0 \circ \phi, I_1\bigr)
\;+\; \lambda \,\mathcal{R}(\phi)
\Bigr].
\end{equation}
Common choices for $\mathcal{D}$ include cross-correlation or mutual information, while $\mathcal{R}$ is often based on the norm of the velocity field or its spatial derivatives.

\subsubsection{VoxelMorph}
 is a deep learning-based framework that leverages a convolutional neural network (CNN), often U-Net style, to predict a dense displacement field $\phi_{\theta}$ directly from the moving and fixed images $(I_0, I_1)$. The network parameters $\theta$ are learned in an unsupervised manner by minimizing a joint loss function:
\begin{equation}
\mathcal{L}(\theta) = \mathcal{D}\bigl(I_0 \circ \phi_{\theta}, I_1\bigr) 
\;+\; \alpha \,\mathcal{R}(\phi_{\theta}),
\end{equation}
where $\mathcal{D}$ measures similarity between the warped moving image $I_0 \circ \phi_{\theta}$ using a local cross-correlation and the fixed image $I_1$, and $\mathcal{R}$ enforces spatial smoothness or regularity of the predicted deformation field, similar to LDDMMs.

\subsubsection{Tracking Methods}

Let $\Omega \subset \mathbb{R}^n$ be the domain of our images.
Each tracking method (see above) provides a discrete velocity field $\mathbf{v}^{(M)}$ with $M$ being any tracking method.
We interpret each $\mathbf{v}^{(M)}$ as describing how points in the image move over time, discretized as:
\begin{equation}
    \phi_{t+1}^{(M)}(\mathbf{x}) = \phi_{t}^{(M)}(\mathbf{x}) + \Delta t \, 
    \mathbf{v}^{(M)}\!\bigl(\phi_{t}^{(M)}(\mathbf{x})\bigr),
    \quad
    \phi_{0}^{(M)}(\mathbf{x}) = \mathbf{x}.
\end{equation}

After iterating these updates from $t = 0$ to $t = T$, we obtain the final displacement field $\phi^{(M)}(\mathbf{x}),$
which maps any point $\mathbf{x} \in \Omega$ to its new location under the chosen velocity field.

Given an input image $I : \Omega \to \mathbb{R}$, we produce the warped image $I'(\mathbf{x})$ by evaluating
\begin{equation}
    I'(\mathbf{x}) = I\bigl(\phi^{(M)}(\mathbf{x})\bigr)
\end{equation}

using the scipy function \verb|map_coordinates|~\cite{scipy_ndimage_map_coordinates}.

\subsection{Analysis Methods}

To assess the accuracy of the tracking algorithms, we manually labeled three important point structures of the patient's head in each of the videos. We labeled the rightmost point of the first vertebra, the leftmost point of the hard palate, and the leftmost point of the mandible to best describe the motion caused by the head and jaw motion. We determined the mean precision error~(MAPE). To quantify the registration efficiency, we measured the mean square error to determine intensity differences and the structural similarity index measure~(SSIM~\cite{ssim}) to account for anatomically close relationships.

\begin{figure}[b!]
    \centering
    \includegraphics[width=0.99\linewidth]{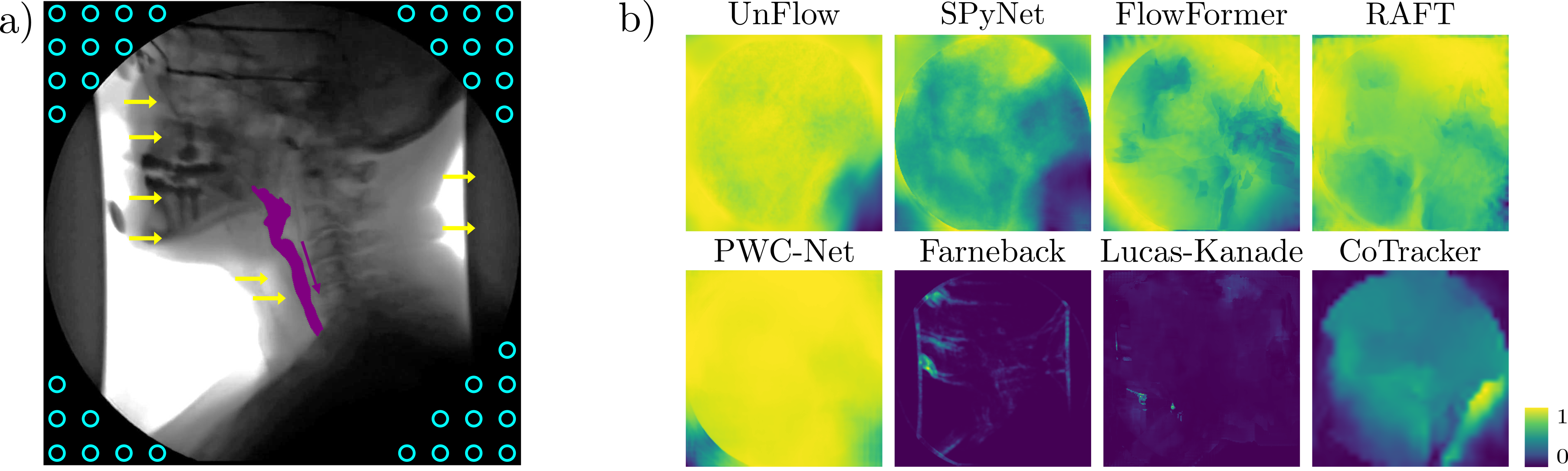}
    \caption{VFSS recording motion sources and their estimation. a) shows a VFSS frame, where the blue circles mark the static background, yellow arrows indicate the moving head, and the bolus with its moving direction is highlighted in purple. b) presents the normalized summed flow fields throughout the same recording as in a) for Deep Learning-based optical flow methods (consisting of UnFlow, SPyNet, Flowformer, RAFT, and PWC-Net), the Farneback and  Lucas-Kanade algorithm as well as one tracking-based algorithm (CoTracker).}
    \label{fig:of_analysis}
\end{figure}

\section{Results}
\begin{figure}[b!]
    \centering
    \includegraphics[width=0.99\linewidth]{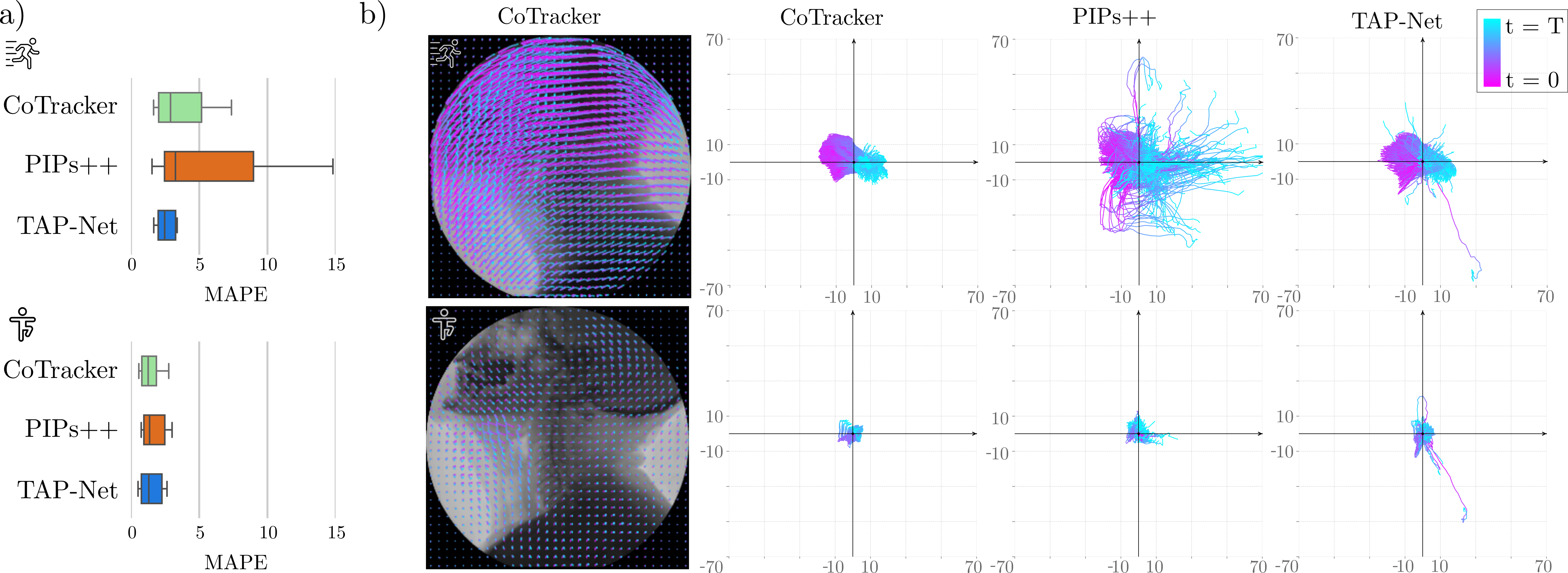}
    \caption{Markerless tracking reliably recovers moving parts. a) MAPE evaluation of the tracking algorithms for three landmarks. b) Visualizes the characteristics of the point tracks with individual trajectories. The magenta-blue color gradient encodes the transition from $t=0$ to $t=T$. The upper and lower panels in a) and b) show the results for images with and without unwanted motion respectively.}
    \label{fig:tracker_analysis}
\end{figure}

The analysis of swallowing physiology presents challenges, as shown in \Cref{fig:of_analysis}a). It includes a static background, bolus transit, and unwanted head movement. To address these complexities, we assessed whether optical flow could effectively estimate different motion sources in VFSS data. However, as shown in \Cref{fig:of_analysis}b), deep learning-based optical flow methods performed poorly, detecting excessive motion throughout the recording. Traditional methods like Farneback and Lucas-Kanade captured only minimal motion aspects. We then explored markerless tracking algorithms for motion pattern detection in medical imaging. Among them, CoTracker demonstrated promising results despite not being specifically trained on or designed for medical imaging data.
We then systematically tested three SOTA markerless tracking algorithms, namely CoTracker, PIPs++, and TAP-Net on manually annotated key landmarks (see Methods). We found only minor differences across tracking algorithms when analyzing the MAPE. In recordings with little motion, the median MAPE is around 1\%, whereas in motion-dominated recordings, it increases to approximately 3\% with larger deviations observed for PIPs++~(\cref{fig:tracker_analysis}a).

Knowing that all trackers are able to track manually selected landmarks, we were interested in an unsupervised approach to capture global motion. We investigated several strategies: random point sampling, good features to track~\cite{gftt}, and a uniform grid on all three markerless tracking methods. We found that the uniformly distributed points best represented the overall patient motion. Therefore, we focus in the following on analyzing the uniform grid (Suppl. Movie 1). 

\begin{figure}[b]
    \centering
    \includegraphics[width=0.99\linewidth]{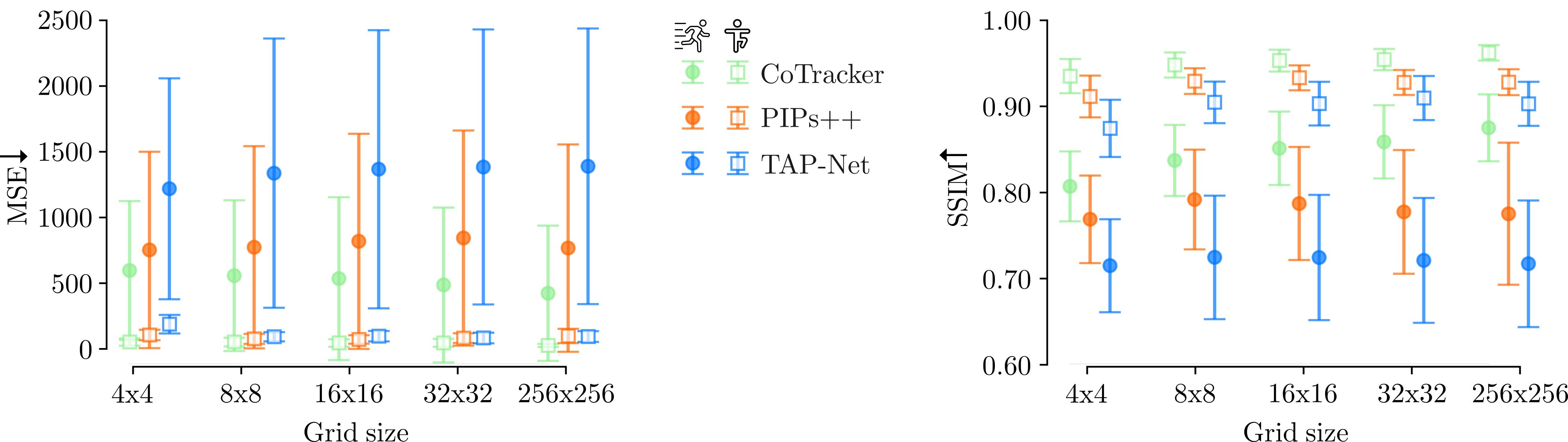}
    \caption{Analysis of the influence of the grid size on the MSE and SSIM \textbf{per video} for the three evaluated tracking algorithms CoTracker, PIPs++, and TAP-Net.}
    \label{fig:gridsize_analysis}
\end{figure}

\Cref{fig:tracker_analysis}b provides an overview of the grid point trajectories for a patient with hardly any motion (lower panel) and one with significantly higher motion (upper panel). 
All three trackers showed a similar angle and path distribution of points over the entire video. CoTracker provides a clear uniform characteristic of point trajectories, whereas PIPs++ and TAP-Net suffer from individual outliers that do not correspond to the general motion present in the video. We consistently observe this behavior for PIPs++, especially in videos with high-motion components. TAP-Net shows this behavior for both high-motion and low-motion videos. In summary, CoTracker is best at capturing the overall motion by showing a very uniform profile across all point trajectories. This confirms our assumption that there is not necessarily the "best point" or ROI to capture the motion; however, the sparse representation of the image with regularly spaced points supports the analysis of head movement.

We hypothesized that we could utilize the regular grids to estimate a displacement field $\Phi$ suitable for a non-linear registration step (see Methods). Therefore, we next analyzed the optimal grid size to estimate the full size $\Phi$. \Cref{fig:gridsize_analysis} shows the MSE and SSIM for each algorithm, averaged across motion and non-motion recordings, against varying grid sizes (see Suppl. Movie 2).
CoTracker performs best with the full, dense grid ($256\times256$), reaching an SSIM of 0.875/0.926 and an MSE of 423.6/26.6 for motion/non-motion recordings, respectively. In contrast, TAP-Net performs the worst among tracker-driven motion correction methods (see~\cref{tab:ssim_mse}, Suppl. Movie 3). Notably, even very small grid sizes (4$\times$4) exhibit stable motion correction capabilities for CoTracker.

\begin{table}[t]
    \centering
    \caption{Comparison of SOTA methods ANTs, VoxelMorph, and LDDMM against our markerless tracker-based registration method with a grid size of $256\times256$ in terms of SSIM and MSE \textbf{per video}. The analysis considers videos with higher patient motion and nearly motionless recordings.(MC = map\_coordinates)}
    \begin{tabular}{c|l|r|r|r|r}
        \hline
        \multirow{2}{*}{} & & \multicolumn{2}{c|}{\includegraphics[width=0.04\textwidth]{fig/motion.png}} 
        & \multicolumn{2}{c}{\includegraphics[width=0.04\textwidth]{fig/arts.png}} \\
        & Method & \multicolumn{1}{c}{SSIM $\uparrow$} & \multicolumn{1}{c|}{MSE $\downarrow$} & \multicolumn{1}{c}{SSIM $\uparrow$} & \multicolumn{1}{c}{MSE $\downarrow$} \\
        \hline
        \hline
        \multirow{3}{*}{\rotatebox[origin=c]{90}{SOTA}} & ANTs & 0.761 ± 0.072 & 1020.4 ± 920.5 & 0.953 ± 0.014 & 27.3 ± 15.2  \\
        & VoxelMorph & 0.822 ± 0.040 & 875.2 ± 865.2 & 0.948 ± 0.012 & 46.8 ± 22.2  \\
        & LDDMM      & 0.850 ± 0.030 & 231.8 ± 299.0 & \textbf{0.962} ± 0.010 & 16.7 ± 6.1  \\
        \hline
        \multirow{3}{*}{\rotatebox[origin=c]{90}{OURS}} & \makebox[1.5cm][l]{PIPs++} $\rightarrow$ MC       & 0.775 ± 0.083 & 767.1 ± 789.6 & 0.928 ± 0.015 & 99.2 ± 54.4   \\
        & \makebox[1.5cm][l]{TAP-Net} $\rightarrow$ MC     & 0.717 ± 0.074 & 1390.2 ± 1048.3 & 0.903 ± 0.026 & 94.8 ± 41.0   \\
        & \makebox[1.5cm][l]{CoTracker} $\rightarrow$ MC    & \textbf{0.875} ± 0.039 & 423.6 ± 514.1 & \textbf{0.962} ± 0.009 & 26.6 ± 11.4  \\
        \hline
    \end{tabular}
    \label{tab:ssim_mse}
\end{table}

\begin{figure}[b!]
    \centering
    \includegraphics[width=0.99\linewidth]{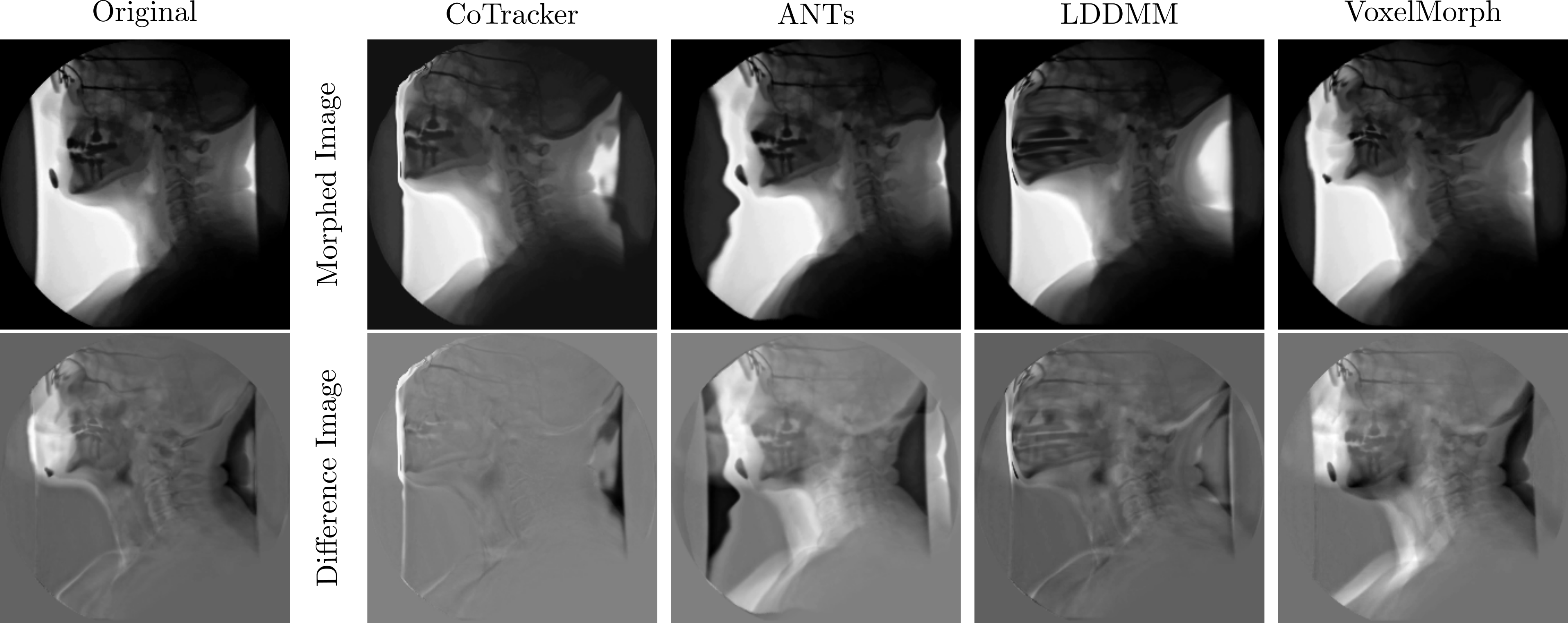}
    \caption{Visual comparison of the last image of the morphed sequence (top row) and the difference between the last and first images (bottom row) for CoTracker, ANTs, LDDMM, and VoxelMorph. Difference image: $ P\in [white > gray=0 > black]$}
    \label{fig:qualitative_comparison}
\end{figure}

Finally, we tested whether our approach is competitive with SOTA registration methods. Qualitatively, the preservation of the patient's anatomy and structure is best with our proposed approach using CoTracker, whereas all competitors, namely VoxelMorph, LDDMM, and ANTs, show high deformation of the patient's head, which influences the patient's appearance and consecutively the unequivocal visibility of the bolus~(\cref{fig:qualitative_comparison}, Suppl. Movie 4). Quantitatively, the CoTracker-based method outperforms LDDMM for motion-dominated recordings but performs comparably for VFSS with no motion, where LDDMM achieves a slightly lower MSE for both types of videos. However, CoTracker surpasses the other two baselines in SSIM and remains competitive in MSE, as shown in~\cref{tab:ssim_mse}. 
Further, our method generalizes to VFSS data from a different hospital which differs greatly in the acquisition settings (see Suppl. Movie 5).

\section{Conclusion}
In our study, we first showed that optical flow methods are not sufficiently usable in the presented context, at least in their current form. We explored an alternative based on markerless tracking algorithms and demonstrated their effectiveness in tracking anatomical structures in VFSS recordings (\cref{fig:tracker_analysis}). Moreover, our results demonstrate that small tracker-based velocity fields are sufficient for motion correction while being computationally efficient (\cref{fig:gridsize_analysis}). Our approach further outperforms competitive baselines quantitatively (\cref{tab:ssim_mse}) and qualitatively (\cref{fig:qualitative_comparison}, Suppl. Movie 4).

We did not explicitly evaluate auxiliary methods to sophistically select points to be tracked, for example, based on segmenting the background, the patient, or anatomical regions.  However, related methods such as VoxelMorph showed that these auxiliary features may be beneficial for registration~\cite{voxelmorph}, suggesting that this strategy could also benefit our proposed framework. 
We further computed the results using raw VFSS recordings without special preprocessing. As this raw medical image footage typically has lower signal-to-noise ratios compared to the footage used in the establishment of tracking algorithms~\cite{cotracker}, denoising algorithms such as Noise2Void~\cite{krull2019noise2void} may have a positive impact on tracking quality.

\bibliographystyle{splncs04}
\bibliography{markerless_motion_correction}

\begin{thebibliography}{10}
\providecommand{\url}[1]{\texttt{#1}}
\providecommand{\urlprefix}{URL }
\providecommand{\doi}[1]{https://doi.org/#1}

\bibitem{cotracker3_facebook}
AI, M.: Cotracker3 - pretrained model weights (2024), \url{https://huggingface.co/facebook/cotracker3/resolve/main/scaled_offline.pth}, accessed: 2025-02-20

\bibitem{voxelmorph}
Balakrishnan, G., Zhao, A., Sabuncu, M., Guttag, J., Dalca, A.V.: Voxelmorph: A learning framework for deformable medical image registration. IEEE TMI: Transactions on Medical Imaging  \textbf{38},  1788--1800 (2019)

\bibitem{lddmm}
Beg, M.F., Miller, M.I., Trouvé, A., Younes, L.: Computing large deformation metric mappings via geodesic flows of diffeomorphisms. International Journal of Computer Vision  \textbf{61}(2),  139–157 (Feb 2005)

\bibitem{bootstapir2024}
Doersch, C., Yang, Y., G{\"o}kay, D., Luc, P., Koppula, S., Gupta, A., Heyward, J., Goroshin, R., Carreira, J., Zisserman, A.: Bootstapir: Bootstrapped training for tracking-any-point (2024), \url{https://github.com/google-deepmind/tapnet}, gitHub repository, Accessed: 2025-02-20

\bibitem{pips2_github}
Harley, A.W.: Pips2: Predicting interactions for pixel streams (2024), \url{https://github.com/aharley/pips2}, gitHub repository, Accessed: 2025-02-20

\bibitem{sam_reg}
Huang, S., Xu, T., Shen, Z., Saeed, S.U., Yan, W., Barratt, D., Hu, Y.: Samreg: Sam-enabled image registration with roi-based correspondence (2024), \url{https://arxiv.org/abs/2410.14083}

\bibitem{kang2010influence}
Kang, B.S., Oh, B.M., Kim, I.S., Chung, S.G., Kim, S.J., Han, T.R.: Influence of aging on movement of the hyoid bone and epiglottis during normal swallowing: a motion analysis. Gerontology  \textbf{56}(5),  474--482 (2010)

\bibitem{cotracker}
Karaev, N., Rocco, I., Graham, B., Neverova, N., Vedaldi, A., Rupprecht, C.: Cotracker: It is better to track together (2024), \url{https://arxiv.org/abs/2307.07635}

\bibitem{hyoidtracking2}
Kim, H.I., Kim, Y., Kim, B., Shin, D.Y., Lee, S.J., Choi, S.I.: Hyoid bone tracking in a videofluoroscopic swallowing study using a deep-learning-based segmentation network. Diagnostics  \textbf{11}(7) (2021). \doi{10.3390/diagnostics11071147}, \url{https://www.mdpi.com/2075-4418/11/7/1147}

\bibitem{hyoidtracking}
Kim, W.S., Zeng, P., Shi, J.Q., Lee, Y., Paik, N.J.: Semi-automatic tracking, smoothing and segmentation of hyoid bone motion from videofluoroscopic swallowing study. PLOS ONE  \textbf{12}(11),  e0188684 (Nov 2017). \doi{10.1371/journal.pone.0188684}, \url{http://dx.doi.org/10.1371/journal.pone.0188684}

\bibitem{krull2019noise2void}
Krull, A., Buchholz, T.O., Jug, F.: Noise2void-learning denoising from single noisy images. In: Proceedings of the IEEE/CVF conference on computer vision and pattern recognition. pp. 2129--2137 (2019)

\bibitem{martin2020best}
Martin-Harris, B., Canon, C.L., Bonilha, H.S., Murray, J., Davidson, K., Lefton-Greif, M.A.: Best practices in modified barium swallow studies. American journal of speech-language pathology  \textbf{29}(2S),  1078--1093 (2020)

\bibitem{aanetworks}
Meng, M., Bi, L., Fulham, M., Feng, D.D., Kim, J.: Enhancing medical image registration via appearance adjustment networks. NeuroImage  \textbf{259},  119444 (2022). \doi{https://doi.org/10.1016/j.neuroimage.2022.119444}, \url{https://www.sciencedirect.com/science/article/pii/S1053811922005614}

\bibitem{cervtracking}
Reinartz, R., Platel, B., Boselie, T., van Mameren, H., van Santbrink, H., ter Haar~Romeny, B.: Cervical vertebrae tracking in video-fluoroscopy using the normalized gradient field. In: Yang, G.Z., Hawkes, D., Rueckert, D., Noble, A., Taylor, C. (eds.) Medical Image Computing and Computer-Assisted Intervention -- MICCAI 2009. pp. 524--531. Springer Berlin Heidelberg, Berlin, Heidelberg (2009)

\bibitem{scipy_ndimage_map_coordinates}
{SciPy Community}: scipy.ndimage.map\_coordinates. \url{https://docs.scipy.org/doc/scipy/reference/generated/scipy.ndimage.map_coordinates.html}, accessed: 2025-02-22

\bibitem{gftt}
Shi, J., Tomasi: Good features to track. In: 1994 Proceedings of IEEE Conference on Computer Vision and Pattern Recognition. pp. 593--600 (1994). \doi{10.1109/CVPR.1994.323794}

\bibitem{lungnodulesegmentation}
Suji, R.J., Bhadouria, S.S., Dhar, J., Godfrey, W.W.: Optical flow methods for lung nodule segmentation on lidc-idri images. Journal of Digital Imaging  \textbf{33}(5),  1306–1324 (Jun 2020). \doi{10.1007/s10278-020-00346-w}, \url{http://dx.doi.org/10.1007/s10278-020-00346-w}

\bibitem{ants}
Tustison, N.J., Cook, P.A., Holbrook, A.J., Johnson, H.J., Muschelli, J., Devenyi, G.A., Duda, J.T., Das, S.R., Cullen, N.C., Gillen, D.L., Yassa, M.A., Stone, J.R., Gee, J.C., Avants, B.B.: The {ANTsX} ecosystem for quantitative biological and medical imaging. Scientific Reports  \textbf{11}(1), ~9068 (Apr 2021). \doi{10.1038/s41598-021-87564-6}, \url{https://doi.org/10.1038/s41598-021-87564-6}

\bibitem{ssim}
Wang, Z., Bovik, A., Sheikh, H., Simoncelli, E.: Image quality assessment: from error visibility to structural similarity. IEEE Transactions on Image Processing  \textbf{13}(4),  600--612 (2004). \doi{10.1109/TIP.2003.819861}

\bibitem{motionsegmentation}
Xue, W., Cao, H., Ma, J., Bai, T., Wang, T., Ni, D.: Improved segmentation of echocardiography with orientation-congruency of optical flow and motion-enhanced segmentation. IEEE Journal of Biomedical and Health Informatics  \textbf{26}(12),  6105--6115 (2022). \doi{10.1109/JBHI.2022.3221429}

\bibitem{us-tracking}
Zhang, Q., Yang, D., Zhu, Y., Liu, Y., Ye, X.: An optimized optical-flow-based method for quantitative tracking of ultrasound-guided right diaphragm deformation. BMC Medical Imaging  \textbf{23}(1) (Aug 2023). \doi{10.1186/s12880-023-01066-7}, \url{http://dx.doi.org/10.1186/s12880-023-01066-7}

\bibitem{pips}
Zheng, Y., Harley, A.W., Shen, B., Wetzstein, G., Guibas, L.J.: Pointodyssey: A large-scale synthetic dataset for long-term point tracking (2023), \url{https://arxiv.org/abs/2307.15055}

\end{thebibliography}
%




\end{document}